\documentclass[letterpaper, 10 pt, conference]{ieeeconf}  

\IEEEoverridecommandlockouts                              

\usepackage{amsmath} 
\usepackage{amssymb}  
\usepackage{booktabs}
\newtheorem{thm}{Theorem}
\newtheorem{lm}{Lemma}

\newtheorem{mydef}{Definition}
\newtheorem{asmpt}{Assumption}

\usepackage{times}
\usepackage{url}
\newcommand{\subparagraph}{}
\usepackage[compact]{titlesec}
\usepackage[font=bf]{caption}
\usepackage{subfig}
\usepackage{xspace}
\usepackage{graphicx}
\usepackage[dvipsnames]{color}
\usepackage{epstopdf}
\newcounter{packednmbr}

\newcommand{\mypara}[1]{\medskip\noindent{\bf#1:}}

\newcommand{\comment}[1]{}

\usepackage{pifont}
\DeclareCaptionType{copyrightbox}

\begin{document}

\title{\LARGE \bf
On the Efficiency and Fairness of Multiplayer HTTP-based \\Adaptive Video Streaming
}

\author{Xiaoqi Yin, Mihovil Bartulovi\'c, Vyas Sekar, Bruno Sinopoli$^{1}$
\thanks{$^{1}$The authors are with the Department of Electrical and Computer Engineering, Carnegie Mellon University,
        Pittsburgh, PA, US. Email: \mbox{yinxiaoqi522@gmail.com}, mbartulo@andrew.cmu.edu, vsekar@andrew.cmu.edu, brunos@ece.cmu.edu}%
}

\maketitle

\begin{abstract}
User-perceived quality-of-experience (QoE) is critical in internet video delivery
systems. Extensive prior work has studied 
the design of client-side bitrate adaptation algorithms to maximize single-player QoE. However, 
multiplayer QoE fairness becomes critical as the growth of video traffic makes it more likely that multiple players
share a bottleneck in the network. Despite several recent proposals, there is still 
a series of open questions. 
In this paper, we bring the problem space to light from a control theory perspective 
by formalizing the multiplayer QoE fairness problem and addressing two key questions in the broader problem space.
First, we derive the sufficient conditions of convergence to steady state QoE fairness under TCP-based bandwidth
sharing scheme. Based on the insight from this analysis that in-network active bandwidth allocation is
needed, we propose a non-linear MPC-based, router-assisted bandwidth allocation algorithm that regards each player as closed-loop systems. We use trace-driven simulation to 
show the improvement over existing approaches. We identify several research directions enabled by 
the control theoretic modeling and envision that control theory can play an important role on guiding 
real system design in adaptive video streaming.
\end{abstract}

\section{Introduction}

In the recent years video streaming became a huge (and still growing) part of the daily internet traffic. In US, Netflix and YouTube alone account for 50\% of download traffic during the peak hours (8pm - 11pm). User-perceived quality-of-experience (QoE) is critical in the Internet video delivery system as it impacts user engagement and revenues of video service providers~\cite{dobrian2011understanding}.

Given that there is little in-network support of QoE in the complex Internet video delivery system, client-side bitrate adaptation algorithms become critical to ensure high user-perceived QoE
by adapting bitrate levels according to network conditions.
A significant amount of research efforts has been focused recently on understanding and designing  better bitrate adaptation algorithms~\cite{yin2014toward,yin2015control,huang2015buffer,li2014probe,sun2016cs2p}.

While client-side bitrate adaptation is critical to ensure high QoE for single player regarding available 
bandwidth as given by a black box, as video traffic becomes predominant on the internet, 
it is more and more likely that multiple video players will share
bottlenecks and compete for bandwidth in the network~\cite{akhshabi2012what, confused}. Such scenarios can be seen in home network, commercial building network, and campus networks,
where multiple devices (e.g., HDTV, tablet, laptop, cell phone, etc.) connect to Internet by a single Wifi router.
In these cases, in addition to single-player QoE, the multi-player QoE fairness becomes a critical issue.

While there have been several practical proposals to address multiplayer QoE fairness problem by designing better player bitrate adaptation algorithms~\cite{conext12,li2014probe}
and network-assisted bandwidth allocation schemes~\cite{Cofano2016Design, Mansy2015Network, Georgopoulos2013Towards}, there are still a lot of open questions in this space. For example, will the interaction among 
different classes of bitrate adaptation algorithms lead to instability? Is centralized, in-network or server-side control necessary to ensure multiplayer QoE fairness?
How to design distributed control schemes with information exchange to achieve QoE fairness?
We envision that this rich and broad problem space presents significant opportunities for control theory to provide insights to a real networking problem and
to guide real system design.

As such, our goal in this paper is to bring the problem space to light from a control theory perspective. 
As a first step in this direction, we 
formalize the multiplayer QoE fairness problem and address a subset of the key questions. 



We start from building a formal mathematical model of the multiplayer joint bandwidth allocation and bitrate adaptation problem,
extending the single-player bitrate adaptation model from prior work~\cite{yin2014toward,yin2015control}. We first focus on the steady-state problem,
and convert the multiplayer fairness problem as the stability analysis of an equilibrium of a non-linear dynamical system.
We derive sufficent conditions under which multiple players with same/different bitrate adaptation policies can 
converge to QoE fairness with TCP-based bandwidth sharing at the bottleneck, and found that 
TCP-based network bandwidth sharing is not sufficient to ensure QoE fairness, confirming the observation of a measurement study~\cite{confused} from a theory aspect.
The result of the analysis calls for active, in-network support for better bandwidth allocation.

Given the recent development of smart routers such as Google OnHub router~\cite{googleonhub} and programmable OpenWrt~\cite{openwrt}, 
we envision that a router-based bandwidth allocation scheme is practical in the near future. 
While recent proposals of router-assisted schemes are based on steady-state utility maximization,
we propose a non-linear MPC-based router-assisted bandwidth allocation algorithm that directly models players as close-loop dynamical systems. 
We evaluate  the proposed strategy using trace-driven simulations and find that the router-assisted control outperforms existing steady-state solutions
in both efficiency and fairness, by adaptively allocating more bandwidth to players which has high resolution 
and insufficient buffer level.

In addition to answering concrete key questions, we hope that this work provides insights into an exciting problem space that has received little attention from
the control community and how control theory can potentially make a significant impact on guiding real system design.

\mypara{Summary of Contributions} The main contribution of this paper is summarized as follows:
\begin{itemize}
\item We bring the multiplayer QoE fairness problem to light from a control theory perspective and provide a 
formal model to reason about existing approaches;
\item We provide theoretical analysis of the convergence of TCP-based bandwidth sharing schemes to QoE fairness;
\item We propose a nonlinear MPC-based router-assisted bandwidth allocation algorithm that outperform existing approaches.
\end{itemize}

The rest of the paper is organized as follows: We begin by sketching the problem space of multiplayer QoE fairness
in Section \ref{sec:bg}. We describe system model and formulate QoE fairness optimization problem in Section \ref{sec:model}.
In Section \ref{sec:analysis} we provide analysis of TCP-based bandwidth sharing policies. We propose
router-based bandwidth allocation in Section \ref{sec:router} and evaluate the algorithm in Section \ref{sec:eval}.
Finally, we conclude the paper with future work in Section \ref{sec:concl}.

\section{Background and Related Work}\label{sec:bg}


In this section, we provide a high-level overview of HTTP-based adaptive video streaming and the multiplayer QoE fairness problem. We then sketch the classes of possible solutions and landscape of prior work, and identify the key questions that call for the use of  control theoretic principles.

\mypara{HTTP-based adaptive video streaming}
Today a lot of video streaming technologies use HTTP-based adaptive video streaming (Apple's HLS, Adobe's HDS, etc.). All these video streaming protocols are standardized under the Dynamic Adaptive Streaming over HTTP or DASH. When using DASH each video is divided into multiple smaller segments or "chunks". Each chunk corresponds to a few seconds of play time and it is encoded at multiple discrete bitrates. This is necessary so that the adaptive video player can switch to a different bitrate if necessary after the chunk was downloaded. 

\begin{figure}[t]
\centering
\includegraphics[width=220pt
]{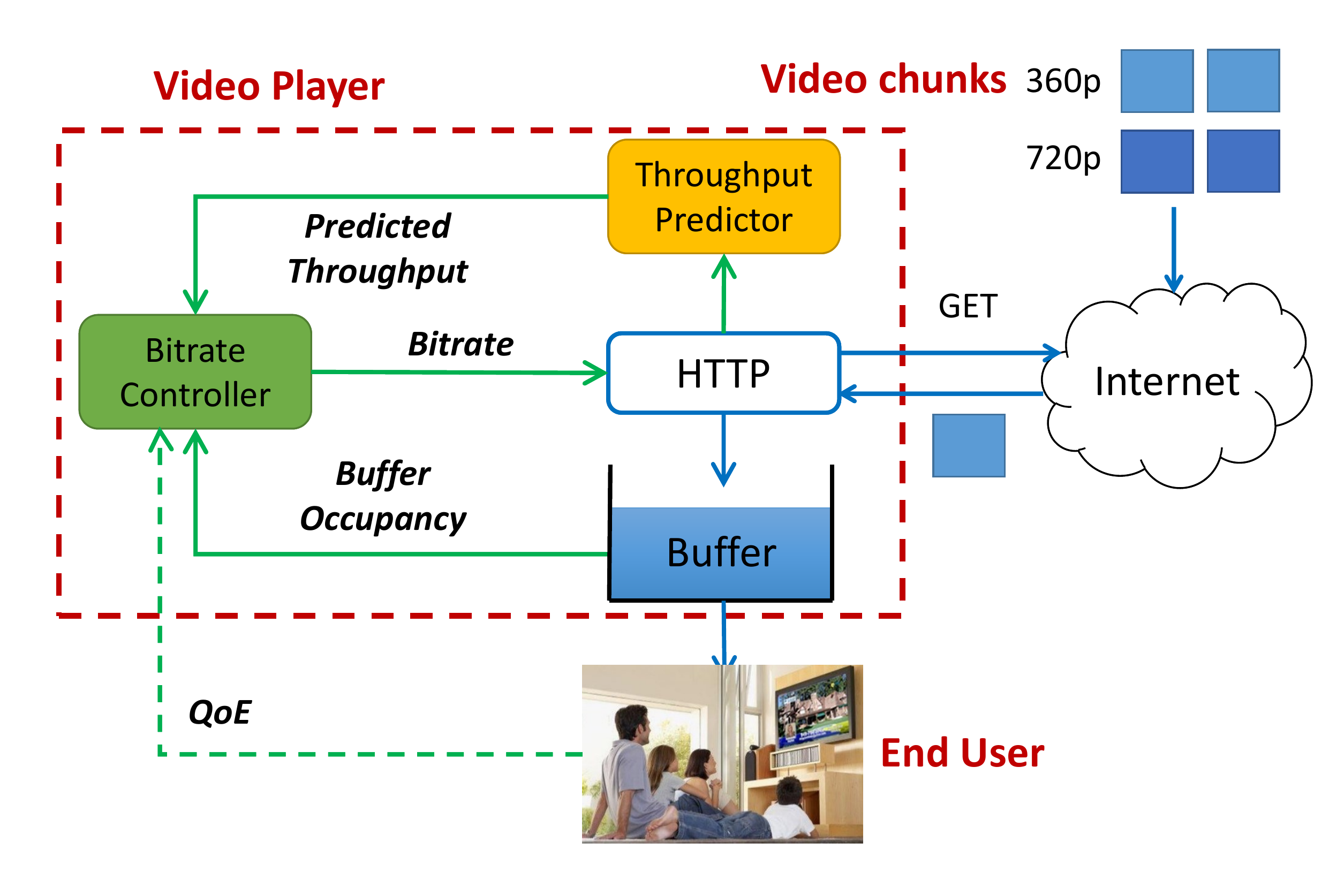}
\vspace{-0.1cm}
\caption{Abstract adaptive video player model}
\vspace{-0.5cm}
\label{fig_mp:player}
\end{figure}

Figure \ref{fig_mp:player} shows an abstract model of an adaptive video player. Video chunks are downloaded via HTTP to a local video buffer, and then played out to users. A \textit{bitrate controller} is responsible to choose the bitrate for each video chunk based on predicted available bandwidth and the state of the buffer, to maximize the user's QoE. A significant amount of work has been focused on the design of the bitrate controller, including rate-based algorithms~\cite{conext12, li2014probe}, buffer-based algorithms~\cite{huang2015buffer, spiteri2016bola}, and hybrid algorithms~\cite{tian2012towards, yin2015control}. In particular, recent work~\cite{yin2015control} provides a control-theoretic framework to understand existing approaches and proposes MPC-based bitrate controller for single-player QoE optimization.

\mypara{Multiplayer QoE fairness} While single-player bitrate adaptation algorithms have been well studied, they consider available bandwidth as a given stochastic variable and maximize QoE for a single player without considering the impact to other players. However, 
When multiple players 
share a bottleneck in the network, 
the efficiency and fairness of QoE across multiple adaptive video players become critical. 

Note that multiplayer QoE fairness includes both fairness in \textit{steady state} and \textit{transient state}. For example, when a HDTV and a tablet share a bandwidth bottleneck in a home network, HDTV should ideally get more bandwidth in steady-state than the tablet as it needs higher-quality video to match the higher resolution. On the other hand, for example, a player with empty buffer is expected to obtain more bandwidth than another with full buffer sharing the same bottleneck, as it needs to quickly accumulate buffer so as to converge quickly to optimal bitrate and avoid rebuffering.

\mypara{Internet video delivery ecosystem} Different from single-player problem, the multiplayer QoE fairness can be affected by a broader range of factors. As such, we zoom out from the adaptive player model in Figure \ref{fig_mp:player} and look at how the internet video delivery ecosystem impacts the multiplayer QoE fairness.

\begin{figure}[t]
\centering
\includegraphics[width=240pt
]{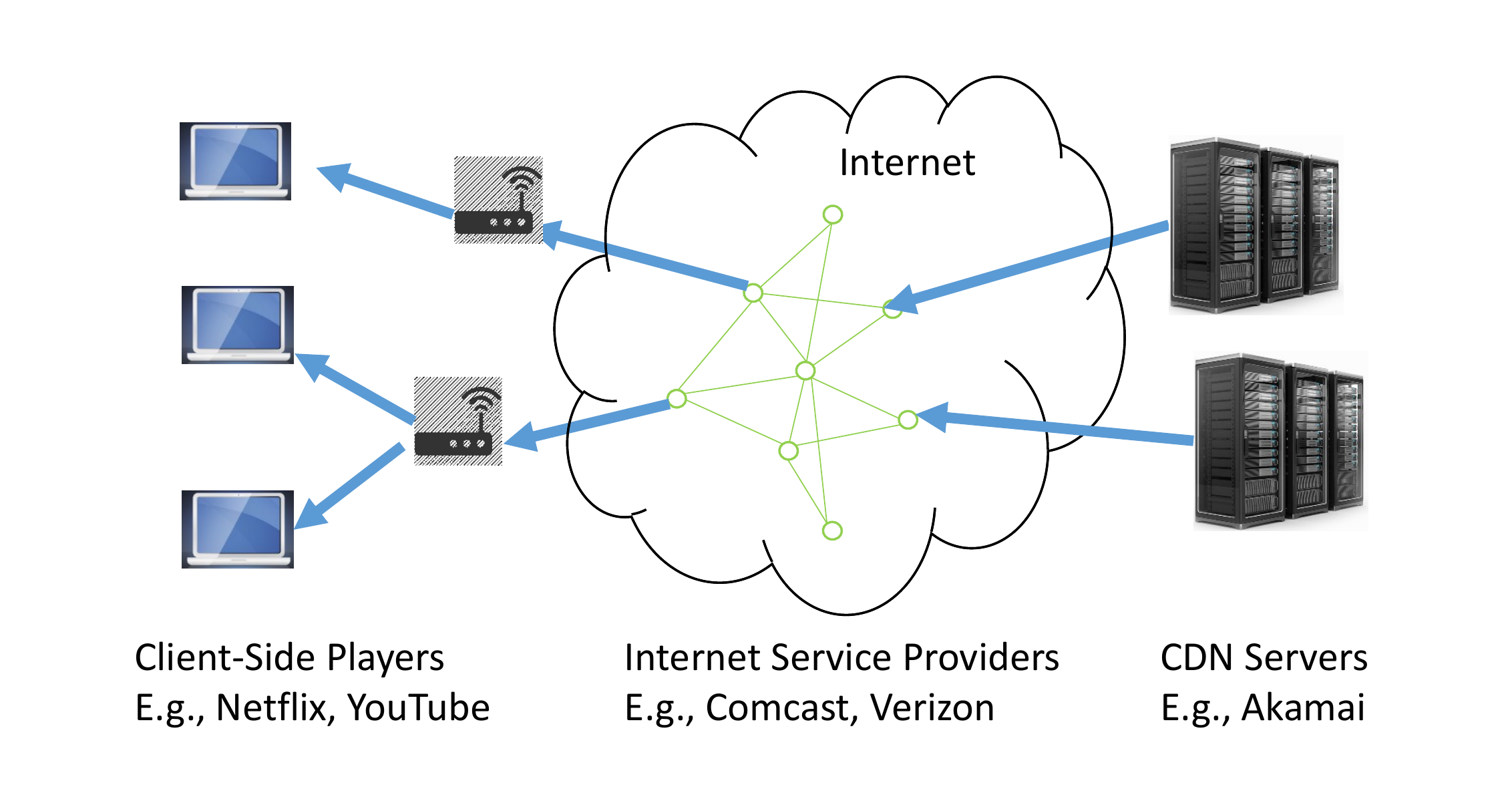}
\vspace{-0.1cm}
\caption{The internet video delivery ecosystem}
\vspace{-0.5cm}
\label{fig_mp:video_ecosystem}
\end{figure}

As shown in Figure \ref{fig_mp:video_ecosystem}, the Internet video delivery ecosystem consists of a variety of entities that has different control capabilities to optimize different objectives. Video source providers, such as Netflix and YouTube,
own the client players and can design client-side bitrate control to optimize the user-perceived QoE; Content delivery networks (CDN), such as Akamai and Level3, place videos in CDN servers at the edge of the internet and assign players to 
best servers in a video session;  
Internet service providers (ISP), such as Comcast and Verizon, control the bandwidth available to CDN servers and client players 
according to agreement
with users; Video quality optimizers, such as Conviva, employ a global view to provide centralized control of bitrate and CDN server selection for client players. 


\mypara{Classes of potential solutions} Given the diverse control capabilities in the internet video delivery system, 
there are several classes of solutions to achieve multiplayer QoE fairness: \textit{player-side}, \textit{in-network}, and \textit{server-side} solutions.

Player-side solutions, such as FESTIVE~\cite{conext12} and PANDA~\cite{li2014probe}, entail designing better bitrate adaptation algorithms for multiplayer QoE fairness. While only requiring player algorithm change and thus easy to deploy, player-side solutions do not alter bandwidth allocation in the network and can suffer from suboptimal bandwidth allocation schemes such as the unideal TCP effect~\cite{confused} and interaction with uncooperative players and cross traffic~\cite{akhshabi2012what}.

In-network solutions, on the other hand, employ active bandwidth allocation in the network to achieve multiplayer QoE fairness. While bottleneck can occur anywhere in the network making such schemes difficult to deploy, there are several recent proposals in particular on router-based bandwidth allocation algorithms to optimize steady-state QoE fairness where router is the single bottleneck shared among players~\cite{Cofano2016Design, Mansy2015Network, Georgopoulos2013Towards}.

Alternatively, server-side solutions regard the server as a single point of control and allocate bandwidth to players~\cite{akhshabi2013server}. However, the actual bandwidth bottleneck can occur in the network instead of server and the computation cost is high when the number of players is too large.

\mypara{Key research questions} The broad problem space for multiplayer QoE fairness has posed a series of key research questions including:
\begin{enumerate}
\item What is the optimal approach and fundamental limitations of each class of solutions?
\item What is the fundamental tradeoff between different classes of solutions?
\item How to design the information exchange scheme to enable coordination of different entities in the video delivery ecosystems to achieve QoE fairness?
\end{enumerate}
As a first step to tackle the broader problem, in this paper we want to develop a principled framework and answer a subset of key questions so as to shed light on the broader problem space and provide useful insights for future work. In the next section, we 
start to develop a formal mathematical model of multiplayer QoE fairness problem.

\section{Modeling}\label{sec:model}

\begin{figure}[t]
\centering
\includegraphics[width=250pt
]{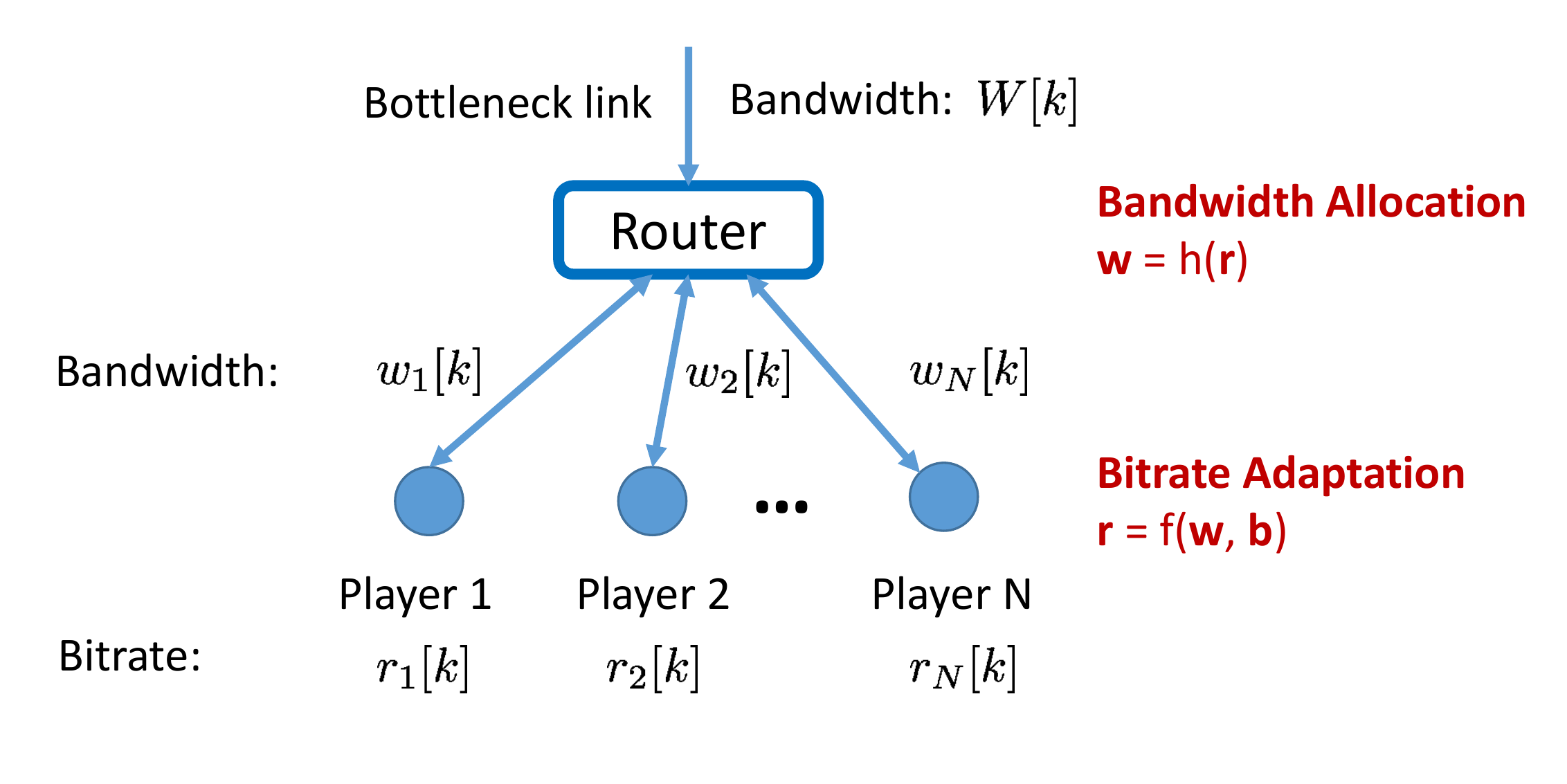}
\vspace{-0.5cm}
\caption{Modeling multiplayer joint bandwidth allocation and bitrate adaptation problem}
\vspace{-0.5cm}
\label{fig_mp:model}
\end{figure}

In this section, we develop a mathematical model for multiplayer HTTP-based adaptive video streaming.
Figure \ref{fig_mp:model} provides an overview of the model.

\mypara{Video streaming model}
We consider a discrete time model with time horizon ${\cal K} = \{1, \cdots, K \}$ with a sampling period $\Delta T$. Let us consider a set of $N$ video players ${\cal P}$ sharing a single bottleneck link with bandwidth $W[k]$ at time $k$. Let $w_i[k]\in\mathbb{R}_+$ be the available bandwidth to the player $i$ at the time $k$, we have: 
\begin{equation}
  \sum_{i\in{\cal P}} w_i[k]\leq W[k], \quad \forall k\in{\cal K}
\end{equation}
We assume this link is the only bottleneck along the Internet path from the video players to the servers. 

Each video player streams video from some video server on the Internet via HTTP. The video is encoded in a set of bitrate levels ${\cal R}$. When downloading video, player $i\in{\cal P}$ is able to choose the bitrate $r_i[k]\in{\cal R}$ of the video at each time step $k$. In constant bitrate encoding,  $r_i\times t$ bits of data need to be downloaded to get the video with $t$ seconds of play time.

Each player has a buffer to store downloaded yet unplayed video. Let $b_i[k]\in[0, \overline{B}_i]$ be the buffer level at the beginning of time step $k$, namely, the amount of play time of the video in the buffer. The buffer accumulates as new video is being downloaded, and drains as video is played out to users. The buffer dynamics of the player $i$ is formulated as follows:
\begin{equation}
  b_i[k+1] = b_i[k]-\Delta T + \frac{w_i[k] \Delta T}{r_i[k]}
\end{equation}

\mypara{QoE objective} The objective of the adaptive video players is to maximize the quality-of-experience (QoE) of users, which is modeled as a linear combination of the following factors: 1) average video quality, 2) average quality change, 3) total rebuffer time and 4) startup delay. For simplicity, in this paper we enforce that there are no rebuffering events, and we only consider the case where all the players have started playback. As such, the QoE utility function $U_i: {\cal R}\times\mathbb{R}_+\times\mathbb{R}_+ \rightarrow\mathbb{R}$ of player $i$ is the formulated as the average QoE of video downloaded over the entire time horizon:
\begin{equation}
U_i = \frac{\sum_{k=1}^K \frac{w_i[k]}{r_i[k]}U_i^P[k]}{\sum_{k=1}^K \frac{w_i[k]}{r_i[k]}} 
\end{equation}
where $U_i^P[k]$ is the QoE of the video downloaded in time $k$: 
\begin{equation}
U_i^P[k] = q_i(r_i[k]) - \mu_i \left| q_i(r_i[k]) - q_i(r_i[k-1]) \right| 
\end{equation}
Note that $q_i:{\cal R} \rightarrow \mathbb{R}$ is the function that maps bitrate to the video quality perceived by users. We assume $q_i(\cdot)$ to be positive, increasing and concave to model the diminishing return property. $\mu_i$ is the parameter that defines the trade-off between high average quality and less quality changes. The larger $\mu_i$ is, the more reluctant the user $i$ is to change the video quality.

\mypara{QoE fairness} Going from single player to multiplayer video streaming, a natural objective function would be the sum of utilities (QoE) of all users, also known as \textit{social welfare} or \textit{efficiency}, i.e., $\sum_{i\in{\cal P}} U_i$. However, in the context of multiplayer video streaming, QoE fairness among players becomes a critical issue as each player usually serves a different user yet they share the same bottleneck resource. As such, we consider the QoE fairness $F(U_1,\cdots, U_N)$ as the objective, where $F: \Pi_{i\in{\cal P}}{\cal U}_i\rightarrow \mathbb{R} $ is a general fairness measure~\cite{Lan2010Axiomatic}. Specifically, we consider a class of fairness measures known as $\alpha$-fairness~\cite{mo2000fair}, where:
\begin{equation}
  F_{\alpha}(\mathbf{U}) =
  \begin{cases}
    \sum_{i\in{\cal P}}\frac{U_i^{1-\alpha}}{1-\alpha} \quad \ \alpha \geq 0, \alpha \neq 1 \\
    \sum_{i\in{\cal P}} \log U_i \quad \alpha = 1
  \end{cases}
\end{equation}
Note that $\alpha$-fairness is a general fairness measure that satisfies axiom 1,2,3,5 from~\cite{Lan2010Axiomatic}. If $\alpha= 1$, $\alpha$-fairness becomes \textit{proportional fairness}; if $\alpha \rightarrow \infty$, it becomes \textit{max-min fairness}.

\mypara{Multiplayer QoE maximization problem} 
Now we are ready to formulate the multiplayer QoE maximization problem where optimal bitrates $(\mathbf{r}[k], k\in{\cal K})$ and bandwidth  $(\mathbf{w}[k], k\in{\cal K}) $ of players are decided to maximize some QoE fairness measure $F(\mathbf{U})$, given the capacity of the bottleneck link, $(W[k], k\in{\cal K})$:
\begin{align}
  \max \quad & F\left( U_1, \cdots, U_N \right) 
  \\
  \mbox{over} \quad & \mathbf{r}[k], \mathbf{w}[k] \quad \mbox{given } W[k], k\in{\cal K} \\
  s.t. \quad & \sum_{i\in{\cal P}} w_i[k] = W[k], \quad  \forall k\in{\cal K} \label{eq:mabr_start}
  \\
  & b_i[k+1] = b_i[k] - \Delta T + \frac{w_i[k] \Delta T}{ r_i[k] }, \\
  & \quad \quad \forall i\in{\cal P}, k=1,\cdots,K \notag \\
  & \underline{B}_i \leq b_i[k] \leq \overline{B}_i, \quad \forall i\in{\cal P}, k\in{\cal K} 
  \label{eq:bufferlimit}\\
  & w_i[k] \geq 0, r_i[k] \in{\cal R}\quad \forall i\in{\cal P}, k\in{\cal K}\label{eq:mabr_end} 
\end{align}


Ideally, a centralized controller can decide both the bitrate $\mathbf{r}$ and the bandwidth $\mathbf{w}$ for all players to achieve QoE fairness, given the complete information of the system. However, the current practice can be interpreted as a distributed way to solve the problem by primal decomposition with no explicit message passing between players and router: Each player $i$ decides the bitrate of itself according to some \textit{bitrate adaptation policy} $r_i[k+1] = f(w_i[k], b_i[k+1])$, while the bottleneck link (conceptually) decides how to allocate available bandwidth according to some \textit{bandwidth allocation policy} $\mathbf{w}[k] = h(\mathbf{r}[k], \mathbf{b}[k])$. The design of optimal distributed solution is to find optimal $(h, f)$ pairs, i.e., $(h^*, f^*)$. Next, we discuss respectively the design of $h$ and $f$. 

\mypara{Bandwidth allocation policies}
Given that the players in ${\cal P}$ shares a bottleneck link with total bandwidth $W[k]$, i.e., $\sum_{i\in{\cal P}}w_i[k] = W[k]$. A \textit{bandwidth allocation} policy $h: \mathbb{R}^n\rightarrow \mathbb{R}^n$ is a function that maps bitrates $\mathbf{r}[k]$ to bandwidth allocation vector $\mathbf{w}[k]$. Let $h_i:\mathbb{R}^n\rightarrow\mathbb{R}$ be the function that maps $\mathbf{r}[k]$ to $w_i[k]$. 
\begin{equation}
  \mathbf{w}[k] = h(\mathbf{r}[k])
\end{equation}

Under ideal TCP,  all players get the equal share of the total bandwidth, i.e., $w_1[k] = \cdots = w_N[k] = W[k]/N$,. However, in practice, TCP is not ideal in the sense that players with larger bitrate gets larger share of the bandwidth due to the discrete effects~\cite{confused}.
We have the following assumptions of the bandwidth allocation function under unideal TCP according to measurement data in \cite{confused}:
\begin{asmpt}
Under non-ideal TCP, the bandwidth allocation policy $h(\cdot)$ has the following properties:
\begin{enumerate}
\item If $r_i=r_j$, $h_i(\mathbf{r}) = h_j(\mathbf{r})$;
\item If $r_i>r_j$, $h_i(\mathbf{r}) > h_j(\mathbf{r})$;
\item $\frac{\partial h_i(\mathbf{r})}{\partial r_i} > 0$, $\frac{\partial h_i(\mathbf{r})}{\partial r_j} < 0$, $i\neq j$;
\item $\lim_{r_i \rightarrow \infty}h_i(\mathbf{r}) < W$, $\lim_{r_i \rightarrow 0}h_i(\mathbf{r}) > 0$;
\item $h(\cdot)$ is symmetric over $\mathbf{r}$ (does not depend on order of players).
\end{enumerate}
\end{asmpt}

\begin{lm}
  The function $h(\cdot)$ has $1+kn$ fixed points, where $k\in\mathbb{N}$.
\end{lm}

\mypara{Bitrate adaptation policies} Bitrate adaptation policy of player $i$, $f_i(\cdot)$, maps available bandwidth $w_i[k]$ and buffer level $b_i[k]$ to bitrate to choose $r_i[k]$ so as to maximize the QoE of the player. Bitrate adaptation policies have been widely studied by both in academia and in industry, and each video streaming service has its own adaptation policy. To make decisions on what bitrate to choose, there are two classes of algorithms: rate-based (RB) or buffer-based (BB) controllers. 

In a rate-based policy $RB(f_i)$, $r_i[k] = f_i(w_i[k-1])$, where $f_i:\mathbb{R}_+\rightarrow{\cal R}$ is an increasing function. We consider a special case $LRB(\alpha)$ where $f_i$ is an affine function $r_i[k] = \alpha w_i[k-1]$. 

In a buffer-based policy $BB(f_i)$, $r_i[k] = f_i(b_i[k])$, where $f_i:\mathbb{R}_+\rightarrow{\cal R}$ is an increasing function. We also consider the special case $LBB(\alpha, \beta)$ of an affine $f$ function $r_i[k] = \alpha b_i[k] + \beta$.

Note that both RB and BB policies can be regarded as heuristic algorithm to maximize QoE which may lead to sub-optimal solution. However, it is still of great interest to study these policies as they are currently widely deployed in the real-world players, such as Netflix or YouTube.

\section{Analysis of Fairness in Steady State}\label{sec:analysis}

\mypara{QoE fairness in the steady state} Note that an interesting special case of the multiplayer problem is when the system is in steady state, where the video quality and bandwidth of all players stay unchanged. Formally, we have the following definition:
\begin{mydef}
  Given fixed total available bandwidth $W$, the multiplayer video streaming system is in steady state $(\mathbf{r}_0, \mathbf{w}_0)$ if for each player $i\in{\cal P}$:
  \begin{enumerate}
  \item Bitrate and bandwidth stay unchanged, i.e., $r_i[k] = r_{0i}$, $w_i[k] = w_{0i}$, $\forall k\in {\cal K}$;
  \item Buffer level is non-decreasing, i.e., $b_i[k+1]\geq b_i[k]$, $\forall k\in{\cal K}$.
  \end{enumerate}
\end{mydef}
Removing the inter-temporal constraints and inter-temporal component in the objective function, we get the multiplayer QoE fairness problem in steady state where optimal solution
is denoted as $(\mathbf{r}_0^*, \mathbf{w}_0^*)$:
\begin{align}
  \max \quad & f\left( q_1(r_1), \cdots, q_N(r_N) \right) 
  \\
  \mbox{over} \quad & \mathbf{r}, \mathbf{w} \quad \mbox{given } W \\
  s.t. \quad & \sum_{i\in{\cal P}} w_i = W, \\
  & r_i\leq w_i, \quad \forall i\in{\cal P} \\
  & w_i \geq 0, r_i \in{\cal R}, \quad \forall i\in{\cal P}
\end{align} 
Note that this problem is convex given that ${\cal R} = [\underline{R}, \overline{R}]$, and in the case that all players share the same $q_i = q$, the optimal solution is $(\mathbf{r}_0^*, \mathbf{w}_0^*): r_{0i}=w_{0i} = W/N$.


\mypara{Fairness of homogeneous RB players} We first consider the simplest case where all players are using the same rate-based algorithms.

\begin{thm}
  If all players adopt $RB(f)$ bitrate adaptation policies, the following statements are true:
  \begin{enumerate}
  \item $(\mathbf{r}_0, \mathbf{w}_0): \ r_{i0} = f\left(\frac{w}{n}\right), w_{i0} = \frac{w}{n}$ is an equilibrium;
  \item If $h\circ f$ is a contractive mapping, $(\mathbf{r}_0, \mathbf{w}_0)$ is globally asymptotically stable;
  \item If $h\circ f$ is a expansive mapping, $(\mathbf{r}_0, \mathbf{w}_0)$ is unstable;
  \end{enumerate}
\end{thm}

\mypara{Fairness of homogeneous BB players} We consider the case where all players adopt the same buffer-based bitrate adaptation policies and have the same QoE functions.
\begin{lm}
  If all players adopts buffer-based bitrate adaptation policy, $(\mathbf{r}_0, \mathbf{w}_0)$ is an equilibrium if and only if: 
  \begin{enumerate}
  \item $\mathbf{r}_0 = \mathbf{w}_0$;
  \item $h(\mathbf{r}_0) = \mathbf{r}_0$.
  \end{enumerate}
\end{lm}

\begin{thm}
  If all players adopts $LBB(\alpha, \beta)$ bitrate adaptation policy, the following statements are true:
  \begin{enumerate}
  \item $(\mathbf{r}_0, \mathbf{w}_0): \ r_{i0} = w_{i0} = \frac{w}{n}$ is an equilibrium;
  \item If $-\frac{1}{n} <\frac{\partial h_i(\mathbf{r}_0)}{\partial r_j}<0$, $\forall i\neq j$, then the equilibrium is locally asymptotically stable;
  \item If $\frac{\partial h_i(\mathbf{r}_0)}{\partial r_j} < -\frac{1}{n}$, $\forall i\neq j$, then the equilibrium is unstable;
  \end{enumerate}
  \label{trm:2}
\end{thm}

Note that comparing results in homogeneous RB and BB players, we found that the convergence of RB players depends on both bandwidth allocation and bitrate adaptation policies, while convergence of BB players only depends on bandwidth allocation functions. The key reason is that, the bitrate decisions of BB players reflects the state of the player, i.e., buffer level, while the bitrate decisions of RB players does not depend on the internal states.

\mypara{Implications on system design} From the analysis we know that, in homogeneous player case, the convergence of BB players only depends on the characteristics of the bandwidth allocation function $h(\cdot)$, while for RB players, the convergence depends on the composite of bandwidth allocation function $h(\cdot)$ and player adaptation algorithms $f(\cdot)$. This has the following key implications that informs the system design: 

First, the analysis confirms that the router-side bandwidth allocation function is critical to the convergence of both RB and BB players. Given that the player adaptation algorithms are designed by potentially different providers and may not be considering multiplayer effect, it could in turn be beneficial to redesign the bandwidth allocation function to ensure convergence with a larger range of player adaptation algorithms. 

Second, the analysis provides a theoretical guide for the design of RB player adaptation algorithms which helps us better understand why existing design works. Given that the convergence depends on both bandwidth allocation and player adaptation, if TCP-based implicit bandwidth allocation is hard to change, we can design better player adaptation algorithms so that $h\circ f$ is contractive. One example of this principle is the design of FESTIVE~\cite{conext12}, where $f(\cdot)$ function is concave to make sure $h\circ f$ is contractive.



\section{NMPC-based Router-Assisted Bandwidth Allocation for QoE Fairness}\label{sec:router}

Despite a fully distributed scheme, the analysis from the previous section has posed the fundamental limitation of 
TCP-based bandwidth allocation scheme: First, not all $h(\cdot)$ lead to convergence to QoE fairness in steady state even if players have the same QoE function $U(\cdot)$
and use the same class of bitrate adaptation policies $f(\cdot)$. Second, it cannot take into account different QoE goals and will not converge to fairness when players employ different classes of bitrate adaptation policies. As such, in order to achieve multiplayer QoE fairness, we want to design better player bitrate adaptation policies $f_i(\cdot)$ and bandwidth allocation policy $h(\cdot)$.

However, it is difficult to deploy/modify bitrate adaptation policies of all video players as they belongs to different and competing video streaming services, e.g., Netflix, YouTube, Amazon Video, etc. Also, controlling the bandwidth from the player side is difficult as the player runs on top of HTTP and cannot change the underlying TCP protocol. Instead, routers are in a good position to collect information of each player and video stream, and can technically control the bandwidth allocation. As smart routers are becoming more and more pervasive in the home entertainment industry (e.g. Google OnHub router), we envision that router-assisted bandwidth allocation scheme is more practical. Overall, we develop a hybrid router-assisted control for fairness: we keep the player adaptation policies $f_i(\cdot)$ unchanged, and design bandwidth allocation policy $h(\cdot)$ to achieve QoE fairness.

As routers have access to all video streams going through, we assume it can get or learn the following information from each player $i\in{\cal P}$: 
1) current states of the player including bitrate $r_i$, buffer level $b_i$, 2) bitrate adaptation policy $f_i(\cdot)$, 3) QoE function $U_i(\cdot)$. 

Given these information, the router-side bandwidth allocation function $h(\cdot)$ is obtained implicitly by solving the following bandwidth allocation
problem 
in a moving horizon manner, regarding each player as a closed-loop system.
\begin{align}
  \max \quad & F\left( U_1, \cdots, U_N \right) 
  \notag\\
  \mbox{over} \quad & \mathbf{w}[k] \quad \mbox{given } W[k], k\in{\cal K} \notag\\
  s.t. \quad & \eqref{eq:mabr_start} - \eqref{eq:mabr_end} \notag\\
  & r_i[k] = f_i(w_i[k-1], b_i[k]), \quad \forall i\in{\cal P}, k\in{\cal K} \notag
\end{align} 
Note that as the dynamics of players are non-linear, the resulting controller is a non-linear MPC-based controller.


\section{Evaluation}\label{sec:eval}

\subsection{Setup}

\mypara{Evaluation framework} We employ a custom Matlab-based simulation framework. 
The duration of each time step is 2s and the simulation framework works in a synchronized manner:
At the beginning of each 2s interval, the states of the player and the network is updated according to 
player dynamics and previously recorded traces. The bitrate and bandwidth decisions are then made simultaneously.
There is no event in between each 2s interval. Note that this is slightly different from the single-player simulation 
in previous section as the player decisions are not synchronized, i.e., the player can change the bitrate at chunk boundaries, which 
may not necessarily be every 2s. We acknowledge this limitation and will
test in real asynchronized settings in future work.

\mypara{Resource allocation schemes} We compare the following algorithms:
\begin{enumerate}
\item \textit{Baseline}: In baseline scheme, the bandwidth controller knows the $q(\cdot)$ function of all players, and the bandwidth is allocated by 
solving steady-state bandwidth allocation problem at the beginning of each time step. 
Given allocated bandwidth, each player then adopts RB or BB adaptation strategies to choose its bitrate.
This scheme has been seen in recent work~\cite{Cofano2016Design, Mansy2015Network, Georgopoulos2013Towards}.
\item \textit{Router}: In router-assisted scheme, the bandwidth controller knows the QoE functions, states (buffer level, bitrate), and bitrate adaptation
strategies of all players. The router-assisted bandwidth controller works in a moving horizon way: 
At the beginning of each time steps, the controller predict the bandwidth in a fixed horizon to the future,
and solve the router-assisted bandwidth allocation problem in the horizon to decide bandwidth allocation. 
We assume the bandwidth is given in the MPC horizon.
\item \textit{Centralized}: The centralized scheme entails calculating the optimal bandwidth allocation
and the bitrate decisions simultaneously by solving the joint optimization problem. We assume the controller knows the 
entire future bandwidth. While less practical, the centralized controller provides us with an upper bound of the performance.
\end{enumerate}

\mypara{Metrics} We evaluate the algorithms using the following performance metrics:
\begin{enumerate}
\item \textit{$\alpha$-fairness}: We adopt $\alpha$-fairness measure as it is widely used in prior work~\cite{Lan2010Axiomatic}. Specifically, we focus
on two special case of $\alpha$-fairness: 1) $\alpha =0$ corresponding to \textit{social welfare, sum of QoE, or efficiency}; 2) $\alpha = 1$ corresponding to
\textit{proportional fairness}. As $\alpha$-fairness can be decomposed into a component corresponding to efficiency and another component corresponding to
fairness measures that does not depend on fairness~\cite{Lan2010Axiomatic}, we also use social welfare and normalized Jain's index as detailed metrics.
\item \textit{Social welfare}: Defined as sum of QoE of all players, i.e., $\sum_{i\in{\cal P}}U_i$.
\item \textit{Normalized Jain's index}: Defined as the Jain's index of normalized QoE, namely, Jain's index of $\mathbf{U}/(\sum_{i\in{\cal P}}U_i)$. 
Jain's index is widely used in prior work to depict QoE fairness of players~\cite{conext12}, it is defined as $J(\mathbf{x}) = (\sum x_i)^2/(n\cdot \sum x_i^2)$.
\end{enumerate}

\mypara{Throughput traces} We use the throughput trace from FCC MBA 2014 project~\cite{fccdataset}.
The dataset has more than 1 million sessions of throughput measurement, each containing 6 measurement of 5-sec average throughput.
For experiment purposes, we concatenate the measurements from the same client IP and server IP, and use the concatenated traces in
the experiment. To avoid trivial cases where the available bandwidth is too high or too low, we only use traces whose average throughput is 
0 to 3Mbps. Also, we multiply the throughput by the number of players in the experiment to eliminate the scaling effect in multiplayer experiments.

\mypara{Player parameters} The time horizon is discretized by $\Delta t = 2s$.
For simplicity, we assume players can choose bitrate in a continuous range [200kbps, 3000kbps].
We set buffer size to be 30s. 
For QoE functions, we set $\mu = 1$ for all players.
For default settings, players has the following video quality function $q(r) = r^p$, we set $p = 0.6$ by default, making
$q(\cdot)$ function concave. Note that this can be non-concave in general, e.g., we could also use the sigmoid-like functions as suggested 
in \cite{Chiang2009Nonconvex}, however, this will make the objective non-convex.
We let RB players adopt $r[k] = 0.8\times w[k-1]$, while
BB players adopt $r[k] = 100\times b[k]$ by default.

\subsection{End-to-End Results}

In this section, we focus on the end-to-end comparison of the algorithms.

\begin{figure}[t]
\centering
\includegraphics[width=200pt
]{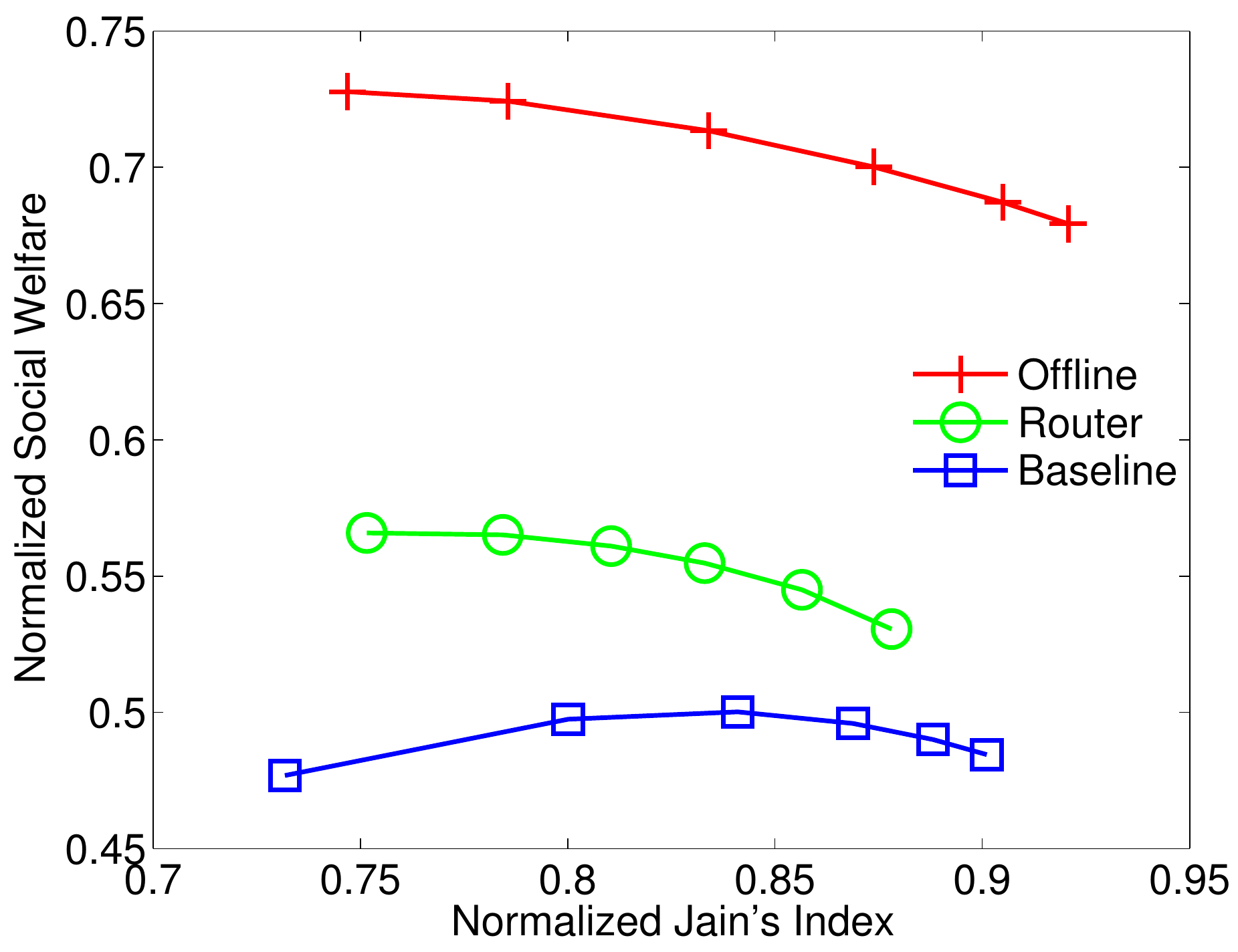}
\caption{Social welfare vs fairness tradeoff}
\vspace{-0.5cm}
\label{fig_mp:alpha_pareto}
\end{figure}

\mypara{Efficiency-vs-fairness tradeoff}
We first evaluate the algorithms in terms of normalized social welfare (sum of QoE) and normalized fairness measure (Jain's index). 
We change $\alpha$ in $\alpha$-fairness in order to get different points on the curve.
Figure \ref{fig_mp:alpha_pareto} shows the pareto front of the algorithms. There are three observations:
First, router-assisted control outperforms baseline controller by 5-7\% in terms of social welfare given the same normalized Jain's index.
For example, if we let normalized Jain's index to be 0.8, router assisted controller achieves ~56\% of optimal, while baseline controller
only achieves 50\% of optimal.
Second, centralized controller significantly outperforms both router-assisted and baseline controller with 15+\% advantage. 
This is because centralized controller has more flexibility on deciding the bitrate for each player, while router-assisted controller
does not have direct control over players' bitrates and can only steer the bitrate by controlling the bandwidth (for RB players) and implicitly buffer level (for BB players).
Third, we observe a natural tradeoff between social welfare and fairness. According to Lan et al.~\cite{Lan2010Axiomatic},
$\alpha$-fairness can be factored into two component: efficiency (social welfare) and fairness measure that satisfies the five axioms and does not depend on scale.
When $\alpha =0$, both centralized and router-assisted controller optimizes social welfare without considering 
the fairness of players. As such, the social welfare at the left most point of the curve is at the maximum. However,
as $\alpha$ is increased, more and more weight is put on the fairness of QoE, leading to increased fairness but
less total QoE. 
Note that this resonates with the observation in prior work~\cite{conext12} on the tradeoff between sum of bitrates and their fairness,
but our proposed algorithms are able to systematically adjust this tradeoff by selecting an appropriate $\alpha$.

\subsection{Sensitivity Analysis}

Next, we conduct sensitivity analysis on a class of key parameters so as to understand the robustness and the reason why router-assisted controller
outperforms existing methods.

\begin{figure}[t]
\centering
\subfloat[]
{
\includegraphics[width=115pt]{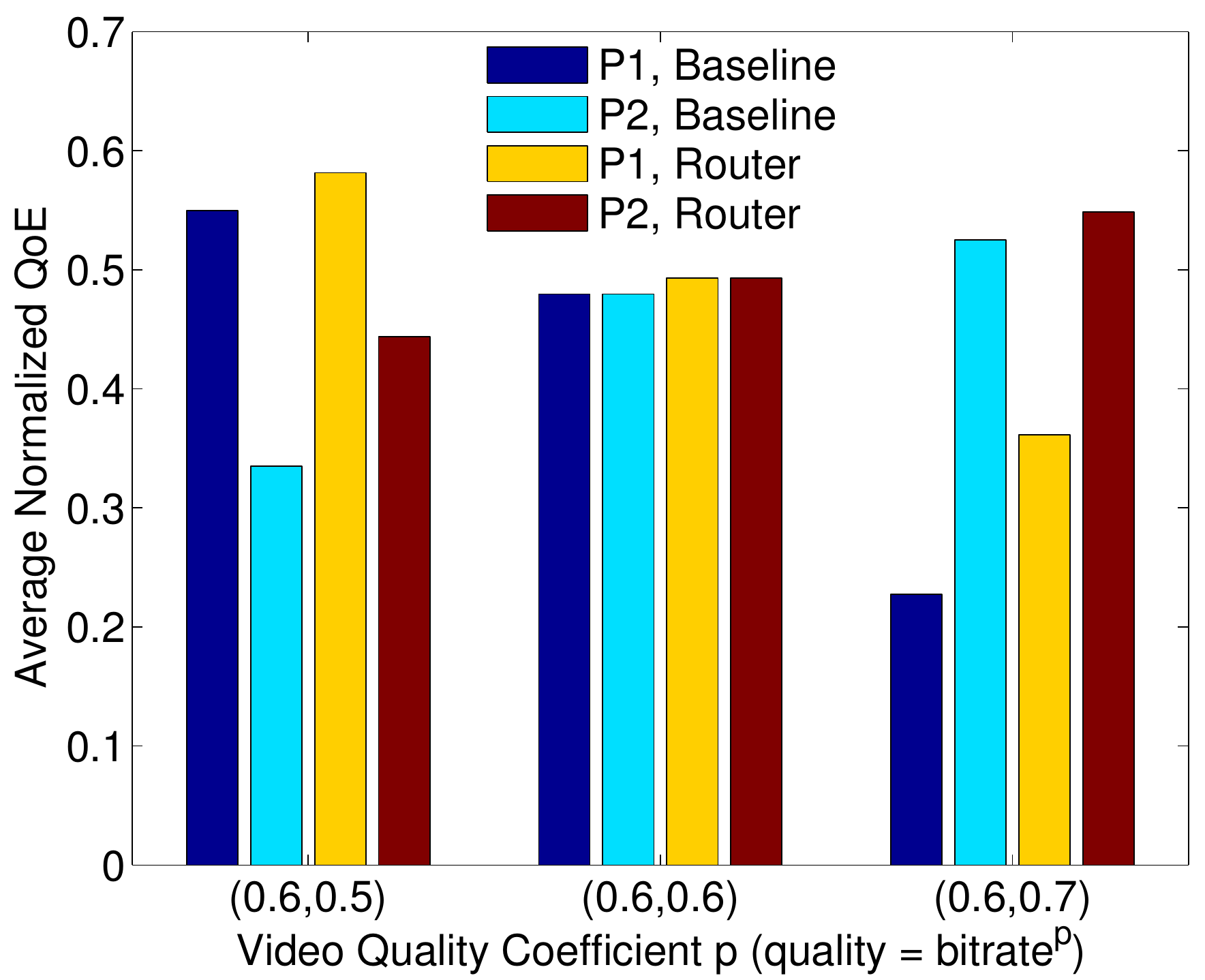}
\label{fig_mp:nqoe_q}
}
\subfloat[]
{
\includegraphics[width=115pt]{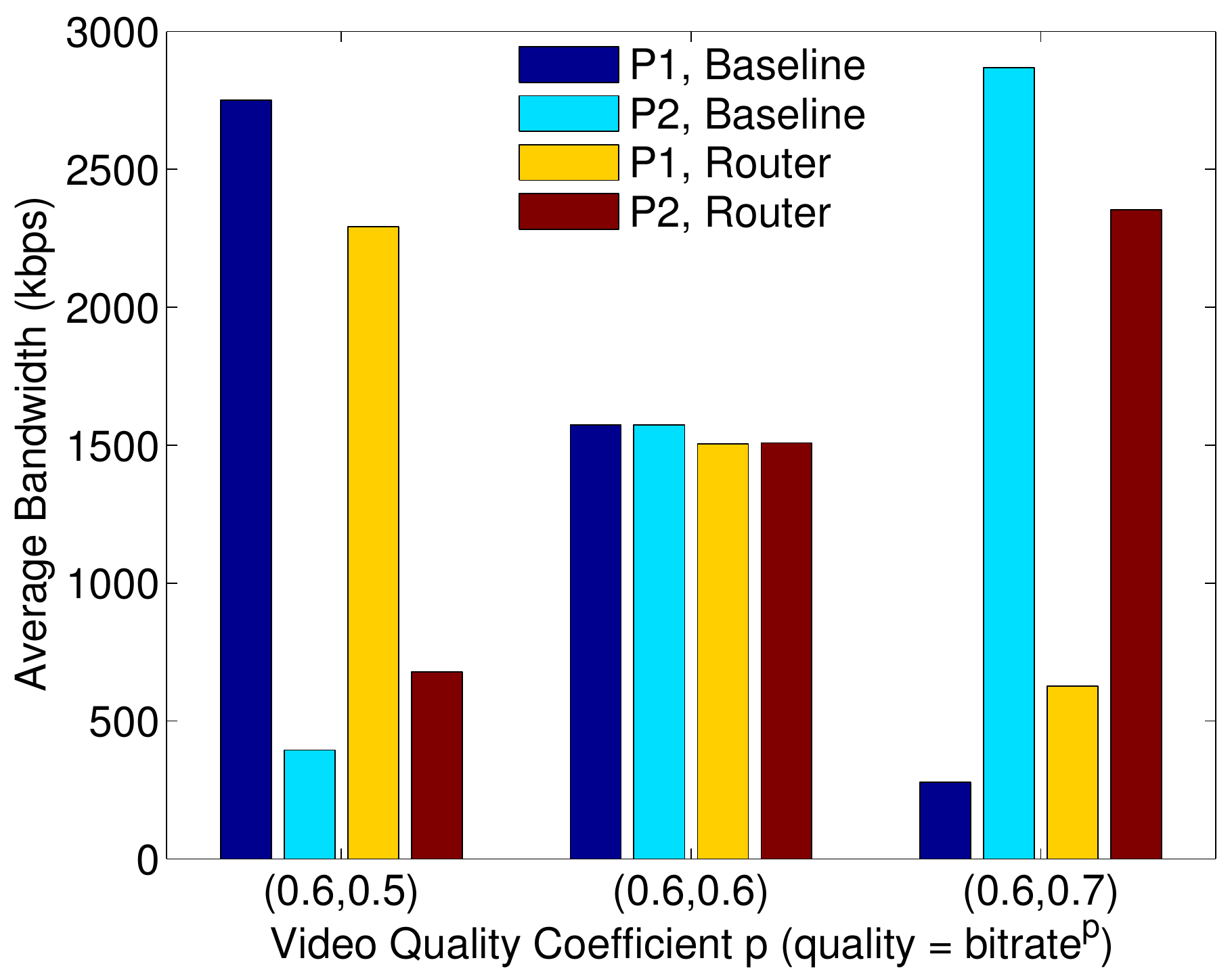}
\label{fig_mp:bw_q}
}
\caption{Impact of QoE functions}
\vspace{-0.5cm}
\end{figure}

\mypara{Impact of QoE functions} We first look at how the algorithms performs under different QoE functions in Figure \ref{fig_mp:nqoe_q}. 
We use two BB players with the same parameters except for video quality functions, i.e., $q(\cdot)$ function. We let $q(r) = r^p$ and vary 
the coefficient $p$. The larger $p$ is, the user-perceived quality is more sensitive w.r.t. bitrate; The smaller $p$ is, the less sensitive the user
is to bitrate. As shown in Figure \ref{fig_mp:bw_q}, both baseline and router assisted controller allocate more bandwidth to the player with larger $p$ and thus requiring 
higher bitrate, as both controllers takes into account the $q(\cdot)$ function in their optimization. 
However, router-assisted algorithm outperforms baseline controllers as it considers player buffer dynamics and lead to faster convergence to optimal bitrates.
In addition, the advantage of router-assisted algorithm over baseline controller is increasing as the video quality coefficients $p$ for different players become
more diverse. Note that this confirms our observation that more bandwidth should be allocated to high-resolution devices in order to achieve QoE fairness.

\begin{figure}[t]
\centering
\subfloat[]
{
\includegraphics[width=115pt]{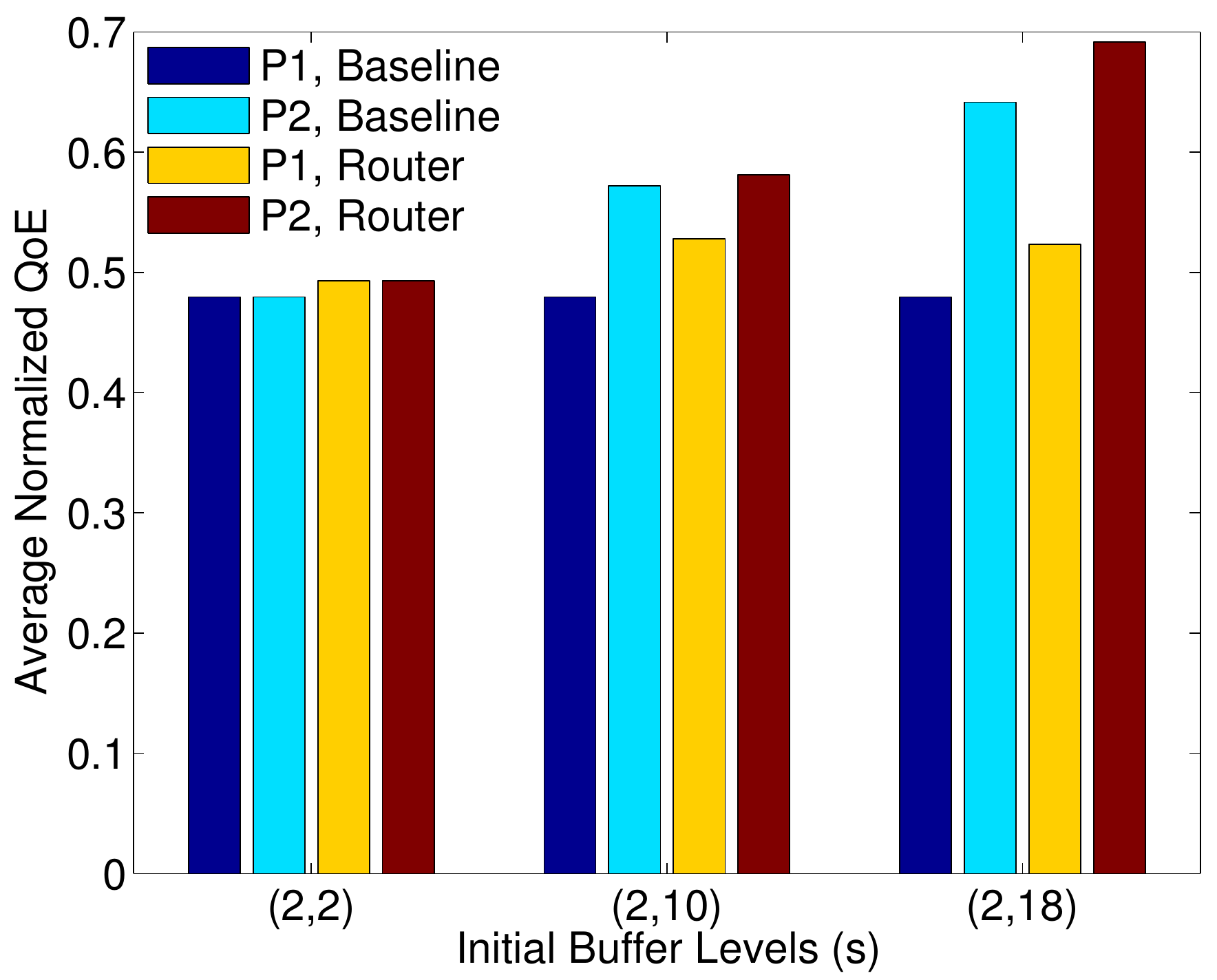}
\label{fig_mp:nqoe_init}
}
\subfloat[]
{
\includegraphics[width=115pt]{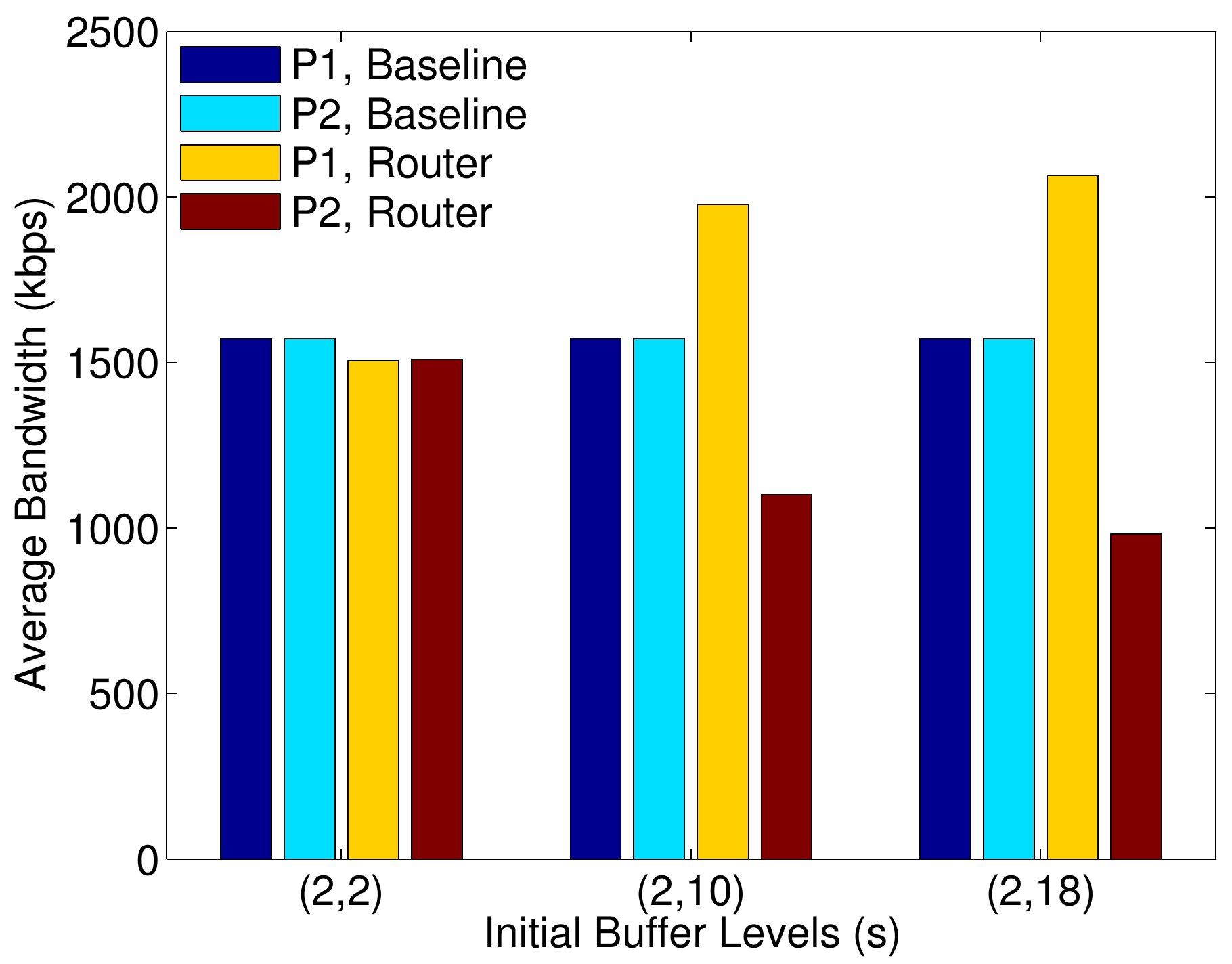}
\label{fig_mp:bw_init}
}
\caption{Impact of initial conditions}
\vspace{-0.5cm}
\end{figure}

\mypara{Impact of initial conditions} We further investigate how the players' initial buffer levels impact the performance. 
Figure \ref{fig_mp:nqoe_init} shows the players' normalized QoE vs different initial conditions, while Figure \ref{fig_mp:bw_init}
shows the bandwidth allocated to players in baseline and router-assisted schemes. There are three key observations:
First, the router-assisted algorithm consistently outperforms baseline solution, increasing the normalized QoE for each player.
Second, the router-assisted algorithm has more advantage over baseline solution when the initial buffer levels for the players become
more diverse. For instance, while  router-assisted and baseline achieves similar performance when both players have 2s buffer initially,
both players' QoE are significantly improved when initial buffer levels are 2s and 18s respectively.
Third, an interesting observation from Figure  \ref{fig_mp:bw_init} is that, while baseline solution does not consider states and dynamics of the players 
and therefore allocate the same bandwidth to both players even one player has much more buffer and need less bandwidth, 
router assisted algorithm allocate less bandwidth to players with full buffer and more bandwidth to player with empty buffer as it needs to quickly accumulate 
buffer so as to stream at high bitrate.
As such, router-assisted algorithm achieves better performance as it takes into account the states and dynamics of the players,
which is critical to players' QoE.

\begin{figure}[t]
\centering
\includegraphics[width=200pt
]{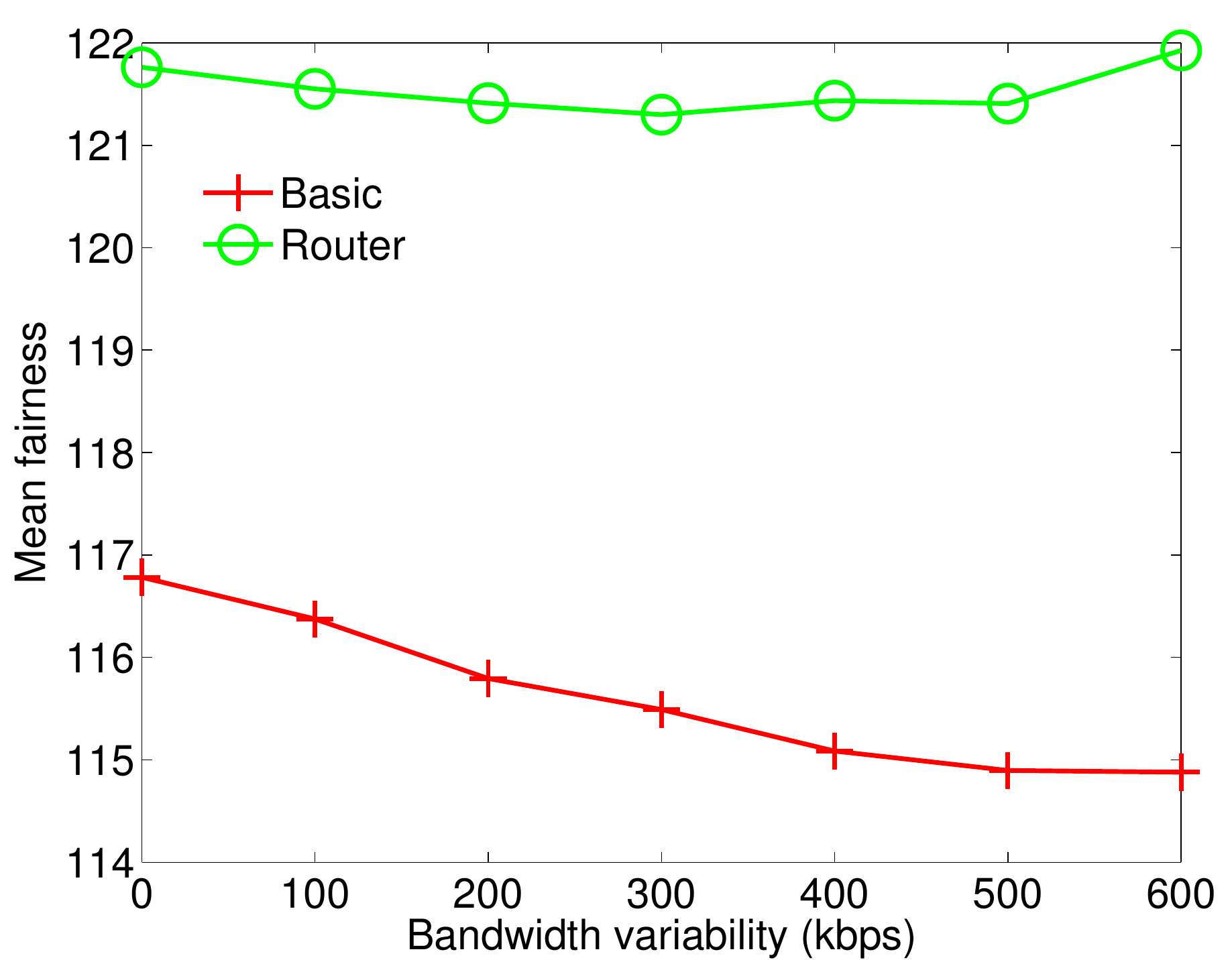}
\caption{Impact of bandwidth variability}
\vspace{-0.5cm}
\label{fig_mp:fairness_variance}
\end{figure}

\mypara{Impact of bandwidth variability} Finally, we investigate how bandwidth variability impacts the performance. 
To showcase that the proposed router-assisted algorithm is more robust to bandwidth variability than the baseline solution, a zero mean Gaussian white noise is added to every bandwidth trace. The variability in bandwidth is increased as we increase the standard deviation of the additive white noise. 
Figure \ref{fig_mp:fairness_variance} shows the mean fairness vs the standard deviation of the additive white noise. Mean fairness is calculated by averaging the results obtained after simulating both algorithms using 100 noisy bandwidth traces.
Furthermore, Figure \ref{fig_mp:fairness_variance} confirms that the router-assisted algorithm is more robust to bandwidth variability as its average fairness stays almost intact while the baseline solution shows a decreasing trend in average fairness as we increase the bandwidth variability. 
This behavior is expected as the router-assisted algorithm uses an adaptive approach to allocate the bottleneck resources leading to better result in highly variable environment.

\subsection{Summary of Results}

Our main findings are summarized as follows:
\begin{enumerate}
\item Given fixed normalized Jain's index, router-assisted algorithm outperforms baseline solution by 5-7\% in terms of social welfare (sum of QoE),
while centralized bandwidth allocation + bitrate control achieves ~70\% of optimal, achieving 15+\% advantage comparing to other solutions.
\item Our sensitivity analysis shows that router-assisted algorithm has more advantage over baseline solution when the QoE functions and initial conditions
of players are more diverse. Moreover, router-assisted algorithm can allocate more bandwidth to players with less buffer while baseline solution fails
to take into account the states of the players.
\end{enumerate}

\section{Conclusion}\label{sec:concl}

Instead of regarding available bandwidth as given by a black box, we further consider the multiplayer interaction
in adaptive video streaming, namely, the joint bandwidth allocation and bitrate adaptation problem in a 
star network. We build a mathematical model and conduct theoretical analysis on the 
convergence of RB/BB players under non ideal TCP assumptions. Given that
convergence is not guaranteed in general, we develop a router-assisted control
which allocate bandwidth to players taking into account their bitrate adaptation
strategies and states. Using trace-drive simulations, we show that our proposed 
router-assisted control outperforms existing QoE-aware bandwidth allocation
algorithms as it can adaptively allocate bandwidth to players with high resolution
and in more urgent need to accumulate buffer.

\bibliographystyle{abbrv}
\bibliography{arxiv}

\end{document}